\documentclass[12pt]{article}
\usepackage{amsfonts, amscd}
\usepackage{amssymb}
\usepackage{eucal,eufrak}
\usepackage{verbatim}
\usepackage{latexsym}
\usepackage{epsfig}

\def\IH{{\mathbb H}}
\def\IR{{\mathbb R}}
\def\IC{{\mathbb C}}
\font\teneufm=eufm10
\font\seveneufm=eufm7
\font\fiveeufm=eufm5
\newfam\eufmfam
\textfont\eufmfam=\teneufm
\scriptfont\eufmfam=\seveneufm
\scriptscriptfont\eufmfam=\fiveeufm

\newcommand{\infinity}{\ensuremath{\infty}}
\newcommand{\dd}{\ensuremath{\partial}}

\newcommand{\fdot}{\partial_t f}

\newcommand{\fddot}{\partial^2_t f}
\newcommand{\drf}{\partial_r f}
\newcommand{\drrf}{\partial^2_r f}

\newcommand{\delr}{\triangle r}
\newcommand{\delt}{\triangle t}

\newcommand{\grad}{\raisebox{.5 ex}{\ensuremath{\bigtriangledown}}}
\newcommand{\lin}{\ensuremath{\langle}}
\newcommand{\rin}{\ensuremath{\rangle}}

\begin{document}
\title{Fast Blow Up in the (4+1)-dimensional Yang Mills Model and the (2+1)-dimensional $S^2$ Sigma Model}
\author{Jean Marie Linhart\\Applied Science Fiction\\8920 Business Park Drive
\\ Austin, TX 78759\\ jlinhart@asf.com}
\date{\today}
\maketitle

\begin{abstract}
We study singularity formation in spherically symmetric solitons of
the (4+1) dimensional Yang Mills model and the charge two sector of
the (2+1) dimensional $S^2$ sigma model, also known as $\IC P^1$ wave
maps, in the adiabatic limit.  These two models are very similar.
Studies are performed numerically on radially symmetric solutions
using an iterative finite differencing scheme.  Predictions for the
evolution toward a singularity are made from an effective Lagrangian and
confirmed numerically. In both models a characterization of the shape
of a time slice $f(r,T)$ with $T$ fixed is provided, and ultimately
yields an new approximate solutions to the differential equations that
becomes exact in the adiabatic limit.
\end{abstract}

\smallskip
\noindent Physics and Astronomy Classification: 02.30.Jr, 02.60.Cb
\medskip

\section{Introduction}

In this paper we study two hyperbolic partial differential equations
that develop singularities in finite time.

The Yang Mills model displays fast blow up, and the $S^2$ sigma model
displays both slow blow up and fast blow up.  In slow blow up, all
relevant speeds go to zero as the singularity is approached.  In fast
blow up the relevant speeds do not go to zero as the singularity is
approached .  The charge 1 sector of the $S^2$ sigma model,
investigated in \cite{Linhart} and \cite{Linhart2}, exhibits
logarithmic slow blow up.  The charge 2 sector of the $S^2$ sigma
model and the similar charge 1 sector of the Yang Mills
(4+1)-dimensional model both exhibit fast blow up.

The Yang Mills Lagrangian in 4 dimensons is a generalization of
Maxwell's equations in a vacuum, and is discussed at length in
\cite{Atiyah}.  We can regard this problem as being that of a motion
of a particle, where we wish our particles to have certain internal
and external symmmetries, which give rise to the various gemoetrical
objects in the problem.  The states of our particles are given by
gauge potentials or connections, denoted $A$, on $\IR^4$, and we identify
$\IR^4$ as $\IH$, the quarternions, $\vec{x} = x_1 + x_2 i + x_3 j +
x_4 k$.  The gauge potentials have values in the Lie albegra of
$SU(2)$ which can be viewed as purely imaginary quarternions,
$\mbox{Im}(\IH)$.  The curvature $F_{ij} = \dd_iA_j - \dd_jA_i +
[A_i,A_j]$, where $[A_i,A_j] = A_iA_j - A_jA_i$ is the bracket in the
Lie Algebra, gives rise to the potential $V(A) = \lin F, F \rin $
which is a nonlinear function of $A$.  The static action is:
\begin{equation} L = \frac{1}{2}
\int \ \lin F_{ij},F_{ij}\rin d^4\vec{x}. \label{YMPLagr}
\end{equation}
The local minima of (\ref{YMPLagr}) are the instantons on 4
dimensional space.  These correspond to solutions of Maxwell's
equations in the vacuum.  We now consider the wave equation generated
by this potential with Lagrangian:
\begin{equation} L = \frac{1}{2}\int \lin\partial_t{A_i},\partial_t{A_i}\rin -\frac{1}{2} \lin F_{ij},F_{ij}\rin d^4\vec{x}. \label{YMLagr}
\end{equation}
Via the calculus of variations, the evolution equation for this (4+1)-dimensional model is
\begin{equation} \partial_t^2{A_i} = -\grad_j F_{ij}.\label{genPDEa}\end{equation}

Theoretical work on the validity of the geodesic or adiabatic limit
approximation for the monopole soltions to the Yang-Mills-Higgs theory
on Minkowski space is presented in \cite{Stuart}.

The two-dimensional $S^2$ sigma model, especially the charge 1 sector,
has been studied extensively over the past few years in \cite{Leese},
\cite{PZ}, \cite{LPZ},\cite{Speight}, \cite{Ward}, \cite{Zakr}.  It is
a good toy model for studying two-dimensional analogues of elementary
particles in the framework of classical field theory.  Elementary
particles are thereby described by classical extended solutions of
this model, called solitons.  This model is extended to (2+1)
dimensions.  The previous solitons are static or time-independent
solutions, and the dynamics of these solitons are studied.  The charge
2 sector of this model is held by general wisdom to be similar to the
(4+1)-dimensional Yang Mills model.

The $S^2$ sigma model can also be regarded as the continuum limit of
an array of Heisensberg ferromagnets.

The static Lagrangian density for the $S^2$ sigma model is given by
\[ L = \int |\grad\vec{\phi}|^2,\]
where $\vec{\phi}$ is a unit vector field.

In the dynamic version of this problem, where
$\phi:\IR^{2+1} \rightarrow S^2$, the Lagrangian is
\[ L = \int_{\IR^2} |\partial_t{\vec{\phi}}|^2 - |\grad\vec{\phi}|^2. \] 
Identifying $S^2 = \IC P^1 = \IC\cup \{\infinity\}$ we can rewrite this as
\begin{equation} L = \int_{\IR^2} \frac{|\partial_t{u}|^2}{(1 + |u|^2)^2} - 
\frac{|\grad u|^2}{(1 + |u|^2)^2}.\label{cp1lag}\end{equation}

The calculus of variations on this Lagrangian yields the following equation of motion for the
$S^2$ sigma model:
\begin{equation}
(1 + |u|^2)(\dd_t^2u - \dd_x^2 u - \dd_y^2 u) = 2\bar{u}(|\dd_t u|^2 - |\dd_x u|^2
- |\dd_y u|^2). \label{genPDEb}\end{equation}
Here  $\bar{u}$ represents the complex conjugate of $u$.

In this paper we  first investigate the charge 1 sector of the
(4+1)-dimensional Yang Mills model and then in the second part go on
to the charge 2 sector of the $S^2$ sigma model.  In each part, we
first derive the evolution equation for our model.  Second, we explain
the numerical scheme used to investigate it.  Third we go over the
predictions made by the geodesic approximation for the model. Fourth,
we go over the results generated from the computer runs.  In both
models the results encompass comparing the trajectory of the evolution
to that predicted by the geodesic approximation, and characterizing
the profile generated at fixed time slices.  In both models our
investigation of the profiles yields an improvement in approximate
solutions of the differential equations, and these approximate
solutions become exact in the adiabatic limit.

\section{The (4+1)-dimensional Yang Mills Model, Charge 1 Sector}

The first things to identify in this problem are the static
solutions to equation (\ref{genPDEa}).  These are simply the 4
dimensional instantons investigated in \cite{Atiyah}.  In \cite{Atiyah} the form
of all such instantons in the degree one sector is shown to be:
\[ A(x) = \frac{1}{2}\left\{\frac{(\bar{x} - \bar{a})dx - d\bar{x} (x -a)}{\lambda^2 + |x-a|^2}\right\}
\qquad x = x_1 + x_2i + x_3 j + x_4 k\in \IH. \] The curvature $F$ of
this potential is computed by $F = dA + [A,A]$ and is:
\[ F = \frac{d\bar{x}\wedge dx}{(\lambda^2 + |x-a|^2)^2}.\]
One notices the denominators are radially symmetric about $x= a$.  

An instanton of unit size centered at the origin would be
\[ A(x) = \frac{1}{2}\left\{\frac{\bar{x} dx - d\bar{x} x}{1 + |x|^2}.\right\}
\]
One motion one can study from these static
instantons would be to consider connections the form
\[A(r, t) = \frac{1}{2}\left\{ \frac{\bar{x}dx - d\bar{x} x}{f(r,t) + r^2}\right\} \qquad r = \sqrt{x_1^2 + x_2^2 + x_3^2 + x_4^2}
\]
and derive an equation of motion for $f(r,t)$ from
(\ref{genPDEa}). This is actually quite computationally extensive.

To get at this, start with connections of the form
\[A(r, t) = \frac{1}{2}g(r,t)\left\{\bar{x}dx - d\bar{x} x\right\}.\]
Using the formula for $F$ given by $F = dA + [A,A]$, and a symbolic algebra program such as Maple, one can use (\ref{genPDEa}) to calculate the differential equation for $g$:

\begin{equation} \partial_t^2{g} = 12 g^2 + \frac{5\partial_r g}{r} + \partial^2_r g - 8 g^3 r^2.\label{protPDEa}\end{equation}

It is much less difficult to now compute a differential equation for $f(r,t)$
if \[ g(r,t) = \frac{1}{f(r,t) + r^2}.\]  It is
\begin{equation} \fddot = \drrf + \frac{5\drf}{r} - \frac{8 \drf r}{f + r^2} 
+ \frac{2}{f + r^2}\left((\fdot)^2 - (\drf )^2\right)
.\label{PDEa}\end{equation}

The static solutions for $f(r,t)$ are simply horizontal lines, $f(r,t)
= c$.  The geodesic approximation for motion under small velocities
states that solutions should progress from line to line, {\it i.e.,\ }
$f(r,t) = c(t)$.  Here the length scale is given by $\sqrt{c}$.
$f(r,t) = 0$ is a singularity of the system, where the instantons are
not well defined.  We observe progression from $f(r,t) = c_0 > 0$ towards
this singularity.  The validity of the geodesic approximation can be
evaluated from how $f(0,t)$ evolves versus how it is predicted to
evolve, and the differences between $f(r,T)$ with $T$ fixed and a
horizontal line.

\subsection{Numerics for the (4+1)-dimensional Yang Mills Model}

In both the (4+1)-dimensional Yang Mills Model and the $S^2$ sigma
model covered in section 3, a finite difference method is used to
compute the evolution of (\ref{PDEa}) and (\ref{PDEc}) numerically.
Unless otherwise noted, centered differences are used consistently, so
that
\begin{eqnarray*} \partial_r g(x) &\approx& \frac{g(x + \delta) - g(x-\delta)}{2\delta}\\
\partial_r^2 g(x) &\approx& \frac{g(x+\delta) + g(x-\delta) - 2 g(x)}{\delta^2}. 
\end{eqnarray*}

In order to avoid serious instabilities in (\ref{PDEa}), the terms
\begin{equation} \drrf + \frac{5\drf }{r}\label{linpart}\end{equation} 
must be modeled in a special way.  We see this revisited for the
$S^2$ sigma model in section 3.  Allow
\[ \drrf + \frac{5\drf }{r} = \mathcal{L}f\]
where \[ \mathcal{L} = r^{-5}\partial_r r^5 \partial_r.\] This
operator has negative real spectrum, hence it is stable.  However the
naive central differencing scheme on (\ref{linpart}) always results in
uncontrolled growth near the origin.  General wisdom holds that when
one has difficulties with the numerics in one part of a problem one
should find a differencing scheme for that specific part in the
natural to that specific part.  Applying this allowed for the removal
of the problem near the origin.  Instead of using centered differences
on $\drrf$ and on $\drf $, we difference the operator:
\[ \mathcal{L}f = r^{-5} \partial_r r^5 \partial_r. \]
The ``natural differencing scheme'' is then \[ \mathcal{L}f \approx
r^{-5}\left[\frac{\left(r + \displaystyle{\frac{\delta}{2}}\right)^{5} \left(\displaystyle{\frac{f(r + \delta) -
f(r)}{\delta}}\right) - \left(r - \displaystyle{\frac{\delta}{2}}\right)^5 \left(\displaystyle{\frac{f(r) - f(r-
\delta)}{\delta}}\right)}{\delta}\right].\]

In \cite{LPZ}, studying the charge 1 sector of the $S^2$ sigma model,
stationary solutions were found to be unstable and to shrink
spontaneously under the numerical scheme.  This model exhibits no such
difficulties.  Unless bumped from the stationary state with an initial
velocity, stationary solutions do not evolve in time.

A discussion of the stability of this numerical scheme is available in
\cite{Linhart}.

With the differencing explained, to derive $f(r,t + \delt)$, one
always has a guess for $f(r,t+\delt)$ given by either the initial velocity,
e.g. $f(r,t+\delt) = f(r,t) + v_0\delt$, or by $f(r,t+\delt) = 2f(r,t)
- f(r,t-\delt)$.  Use this to compute $\partial_t{f}(r,t)$ on the right hand
side of (\ref{PDEa}).  Then solve for $f(r,t +\delt)$ in the
difference for $\fddot(r,t)$, and iterate this procedure to get a
more precise answer.  So, one iterates
\begin{eqnarray*} f(r,t+\delt) &=& 2f(r,t) - f(r,t-\delt) + (\delt)^2\left[
\drrf(r,t) - \frac{5\drf (r,t)}{r}\right.\\ & & - \left.
\frac{2\partial_t{f}(r,t)^2}{f(r,t) + r^2} - \frac{2\drf (r,t)^2}{f(r,t) + r^2}
- \frac{8 \drf (r,t)r}{f(r,t) + r^2}\right],\end{eqnarray*} where all
derivatives on the right hand side are represented by the appropriate
differences.

There remains the question of the boundary conditions.  The function
$f$ is only modeled out to a value $r = R \gg 0$.  Initial data for
$f(r,0)$ was originally a horizontal line.  The corresponding boundary
conditions are that $f(R,t) = f(R-\delr, t)$, and that \[ f(0,t) =
\frac{4}{3} f(\delr, t) - \frac{1}{3} f(2\delr,t)\] {\it i.e.,\ } that $f$ is
an even function.

Subsequent investigation of the model indicated that the appropriate
form for $f(r,t)$ was a parabola instead of a line, and the $f(R,t)$
boundary condition was changed to reflect this. For the runs with
parabolic initial data, we set the boundary condition at $R$ to be:
\[\drf (R,t) = \drf (R-\delr, t)\frac{R}{R-\delr}.\]

\subsection{Predictions of the Geodesic Approximation for the 
(4+1)-dimensional Yang Mills Model}

Equation (\ref{YMLagr}) gives us the Lagrangian for the general
version of this problem.  We are using
\[ A = \frac{1}{2} \left\{ \frac{\bar{x} dx - d\bar{x} x}{f + r^2}\right\}
\qquad r = \sqrt{x_1^2 + x_2^2 + x_3^2 + x_4^2},\] and we use the
geodesic approximation which predicts we  move close to the moduli space
of these solutions.  Restricting the Lagrangian to this moduli space
gives us an effective Lagrangian.  The portion of the integral given
by
\[  -\frac{1}{4} \int_{\IR^4} \lin F_{ij},F_{ij}\rin d\vec{x} \]
represents the potential energy and integrates to a topological constant,
hence it may be ignored.  We see this same phenomenon again
in section 3.2 in the study of the $S^2$ sigma model.  We need to calculate
\[ \frac{1}{2} \int_{\IR^4} \lin\partial_t{A_i},\partial_t{A_i}\rin d\vec{x}. \]
First calculate
\[ \lin \partial_t{A_i},\partial_t{A_i}\rin = \frac{3r^2 (\fdot)^2}{(f + r^2)^4}.\]
So the Lagrangian is
\[ \int_{\IR^4} \frac{3r^2 (\fdot)^2}{(f + r^2)^4} d\vec{x}, \]
If we take $f = \mbox{constant}$, using $\vec{y} = \vec{x}/\sqrt{f}$
we can rewrite this integral as
\[ \frac{3(\fdot)^2}{f} \int_{\IR^4} \frac{|y|^2}{(1 + |y|^2)^4} d\vec{y}.\]
The integral with respect to $\vec{y}$ converges, hence we have the effective
Lagrangian
\[ L = c\frac{(\fdot)^2}{f}. \]
This is purely kinetic energy.  Since the potential energy is
constant, so is the kinetic energy.  We have
\[ \frac{(\fdot)^2}{f} = k. \]
Integrating this we get
\[ f = (c_1 t + c_2)^2.\]
If $f = 0$ occurs at time $T$, we find
\[ T = -\frac{c_2}{c_1} \]
hence we rewrite this as
\[ f = a(t - T)^2.\]
This is how we predict that $f(0,t)$ will evolve.  This same evolution is
predicted in section 3.2.

\subsection{Results for the (4+1)-Dimensional Yang Mills Model, Evolution of $f(0,t)$}

The computer model was  run under the condition that $f(r,0)
= f_0$ with various small velocities.  The initial velocity is
$\partial_t{f}(r,0) = v_0$, other input parameters are $R = r_{max}$,
$\delr$ and $\delt$.  

The first question to ask is how does the evolution of the origin
occur.  We note that as $r\rightarrow \infinity$ equation (\ref{PDEa}) becomes the regular linear
wave equation 
\[ \fddot = \drrf,\]
and so the interesting nonlinear behavior
is at the origin.  Consequently we track the evolution of $f(0,t)$.

We find that the evolution of $f(0,t)$ evolves exactly as predicted,
as a parabola of the form:
\[f(0,t) = a(t-T)^2.\] 
Unsurprisingly, we obtain this same result in section 3.3 for the
$S^2$ sigma model.

The time to ``blow up'' is the parameter $T$ in this equation.  Recall
that $f_0 = f(r,0)$ is the initial height and $\partial_t{f}_0(r,0) = v_0$
is the initial velocity.  Using a least squares parabolic fit to the
origin data obtained after $f(0,t) \leq 0.5f_0$, one obtains the
parameters $a$ and $T$ for a given origin curve.  Table \ref{op4p1}
shows the behavior.

\begin{table}[H]
\begin{center}
\caption{Parabolic fit to $f(0,t)$ vs. Initial conditions $f_0$ and $v_0$}
\label{op4p1}
\[ \begin{array}{*{4}{r}} f_0 & v_0\  & a\ \ \ \ \  & T\  \\ 
1.0&-0.010&0.00002501&200.1\\
2.0& -0.010&0.00001257&399.4\\
0.5&-0.010&0.00005166&99.0\\
4.0&-0.010&0.00000626&799.7\\
4.0&-0.020&0.00002503&400.1\\
4.0&-0.005&0.00000157&1599.3 \end{array}\]
\end{center}
\end{table}

We calculate the parameters $a$ and $T$ from the initial conditions,
and we find
\[ a  = \frac{v_0^2}{4f_0},\] and 
\[ T = \frac{2f_0}{|v_0|}.\]

\begin{figure}[H]
\begin{center}
\epsfig{file=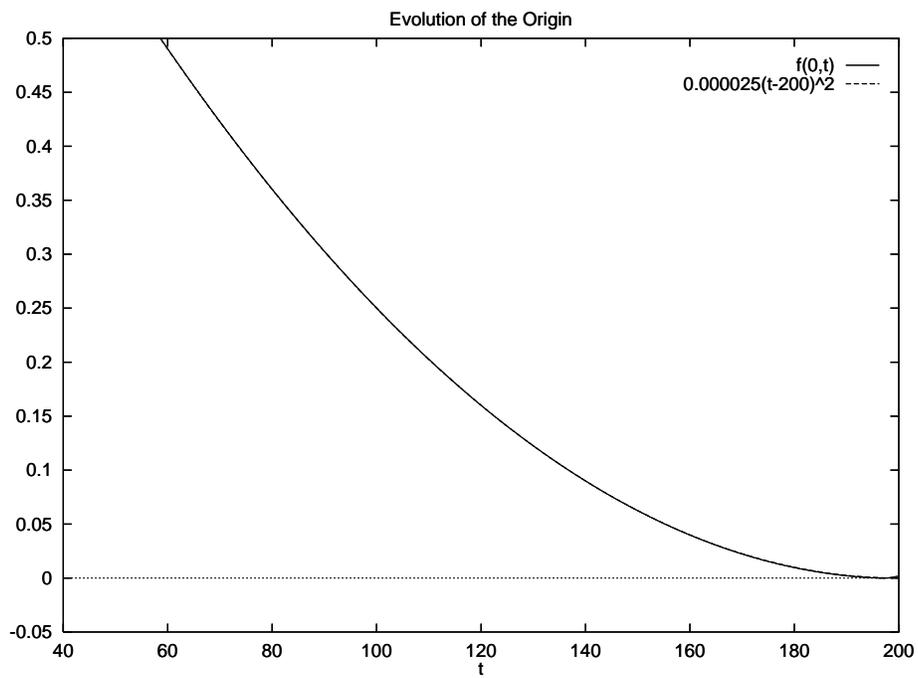}
\caption{Evolution $f(0,t)$.}
\label{o4p1}
\end{center}
\end{figure}

A typical evolution of $f(0,t)$ is given in Figure \ref{o4p1}.  In
this figure, the equation $0.000025(t-200)^2$ neatly overlays the
graph of $f(0,t)$.  This picture represents the evolution where $f_0 =
1.0$ and $v_0 = -0.01$.  Hence $a = \frac{(0.01)^2}{4(1.0)} =
0.000025$ and $T = \frac{2(1.0)}{0.01} = 200,$ as predicted.

\clearpage

\subsection{Characterization of Time Slices $f(r,T)$: Evolution of a Horizontal Line in the (4+1)-dimensional Yang Mills Model}

\begin{figure}[H]
\begin{center}
\epsfig{file=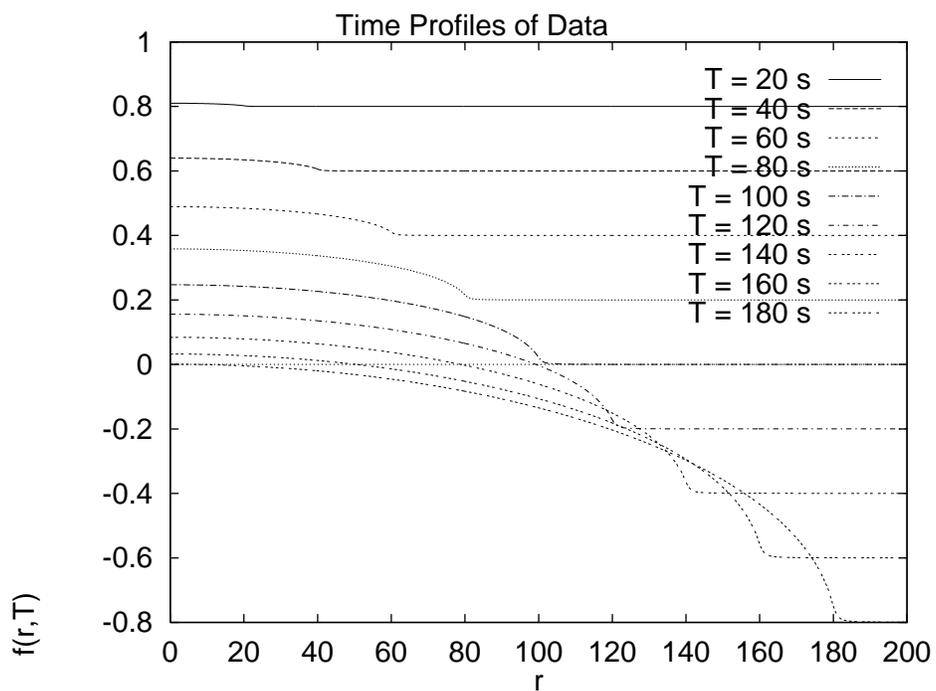}
\caption{(4+1)-dimensional Yang Mills model, Time Slices $f(r,T)$ evolve an elliptical bump at the origin}
\label{tsl4p1}
\end{center}
\end{figure}

The most striking immediate result is that the initial line, $f(r,0) =
f_0,$ evolved an elliptical bump at the origin that grew as time
passed.  This happens again in section 3.4 with the $S^2$ sigma
model.  Figure \ref{tsl4p1} shows this behavior.

The elliptical bumps can be modeled as 
\begin{equation} \frac{x^2}{a^2} + \frac{(y-k)^2}{b^2} = 1.\label{ellform}\end{equation}
The question naturally arises as to how the parameters $a$, $b$ and
$k$ evolve.  This is straightforward:
\[ a = t \]
\[ b = \frac{v_0^2}{4f_0} t^2\]
\[ k = f_0 + v_0t \]

\clearpage

\subsection{Characterization of Time Slices $f(r,T)$: Evolution of a Parabola
in the (4+1)-dimensional Yang Mills Model}

\medskip
The elliptical bump that formed in the evolution of a horizontal line
and the various configurations that ensued after it bounced off the
$r=0$ and $r = R$ boundary suggested that perhaps the curve was trying
to obtain the shape of a parabola.  After all, near $r = 0$, ellipses
are excellent approximations for parabolas of the form

\begin{equation} f(r,t) = pr^2 + h.\label{parabform}\end{equation}

To get the parabola, calculate from the general form of our ellipse
in (\ref{ellform}) 
\[ \frac{dy}{dx} = -\frac{x^2b^2}{(y-k)a^2}\]
so
\[ \frac{d^2y}{dx} = \frac{-b^2}{(y-k)a^2} - \frac{xb^2}{(y-k)^2a^2}\frac{dy}{dx}\]
At $x = 0$, $y-k = b$ and this gives
\[ \frac{d^2y}{dx} = \frac{-b}{a^2}.\]
Recall from the previous section that $b = ct^2$ and $a = t$, so this
gives \[ \frac{d^2y}{dx^2} = -c.\] The identification of $c$ gives \[
\frac{d^2y}{dx^2} = -\frac{v_0^2}{4f_0}.\] So \[p =
-\frac{1}{2}\frac{d^2y}{dx^2} = -\frac{v_0^2}{8f_0}\] 

\begin{figure}[H]
\begin{center}
\epsfig{file=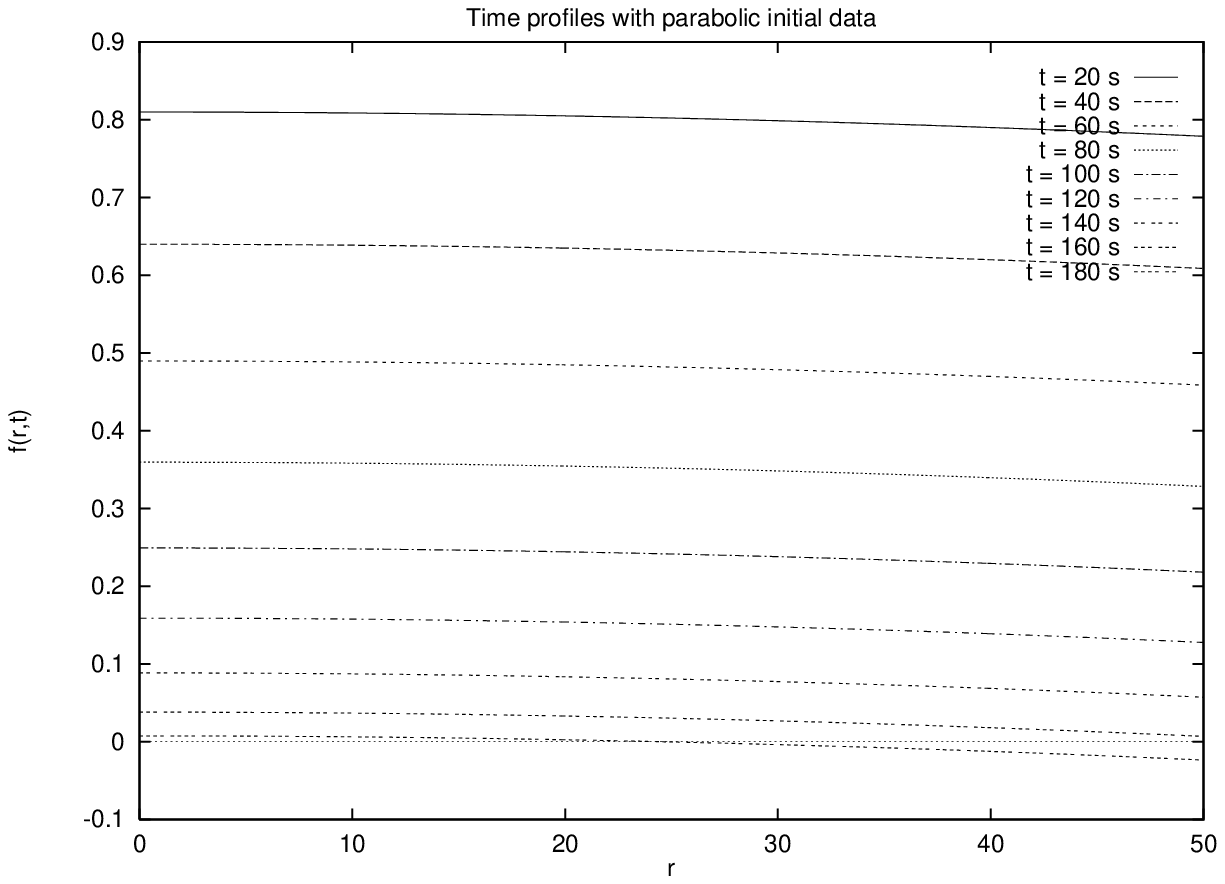}
\caption{(4+1)-dimensional Yang Mills model: Time slices of the evolution of a parabola
are parabolas.}
\label{p4p1}
\end{center}
\end{figure}

When a run is started with this initial data, $\partial_t{f_0} = v_0 =
-0.01$, $f_0 = f(0,0) = 1.0$ and $p = -\frac{v_0^2}{8f_0} =
-0.0000125$, the time slices of the data have this same profile.  This
is shown in figure \ref{p4p1}.

\begin{figure}[H]
\begin{center}
\epsfig{file=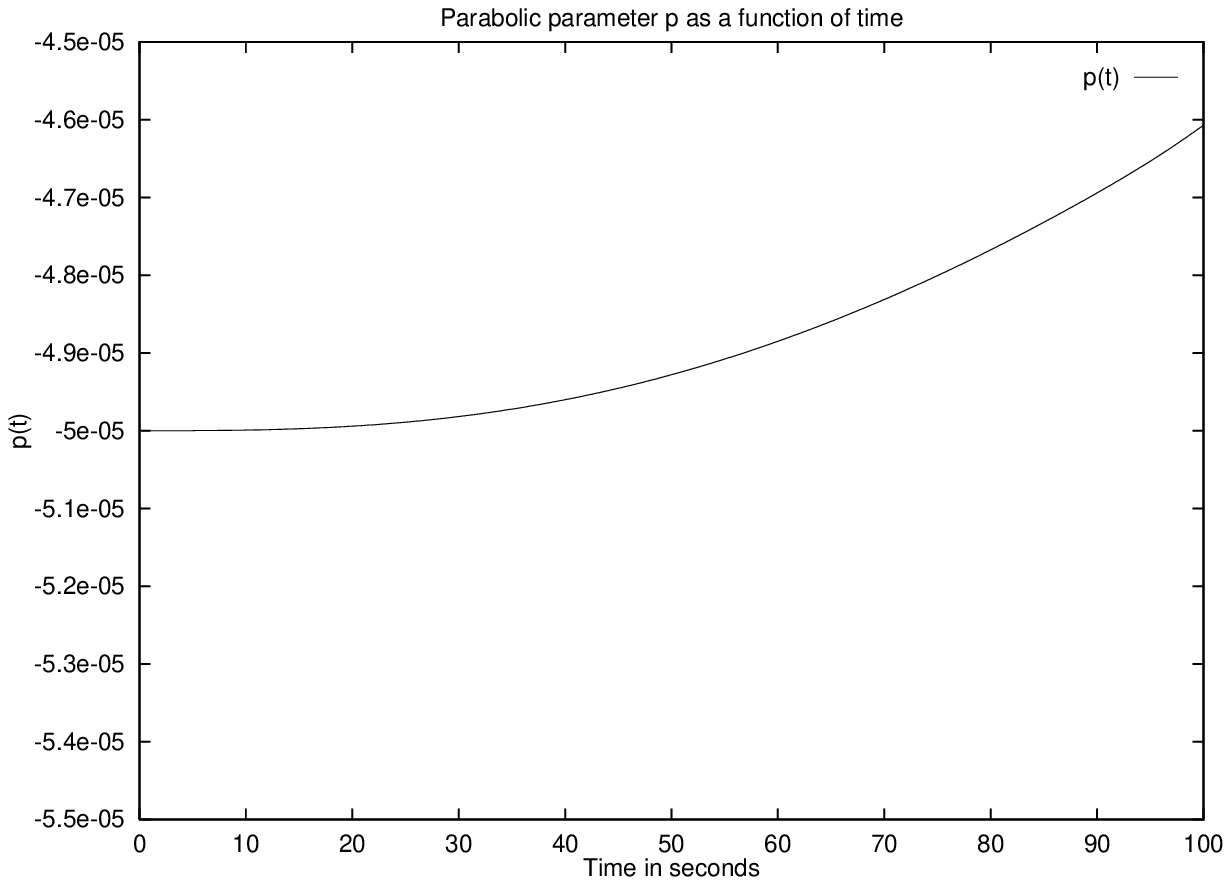}
\caption{(4+1)-dimensional Yang Mills model: evolution of parabolic parameter $p$ with time.}
\label{pp4p1}
\end{center}
\end{figure}

The curvature of the parabola at the origin, as measured by the
parameter $p$ from equation (\ref{parabform}) changes by less than 1
part in 100 during the course of this evolution, a graph of $p$ over
time can be seen in figure \ref{pp4p1}.

\begin{figure}[H]
\begin{center}
\epsfig{file=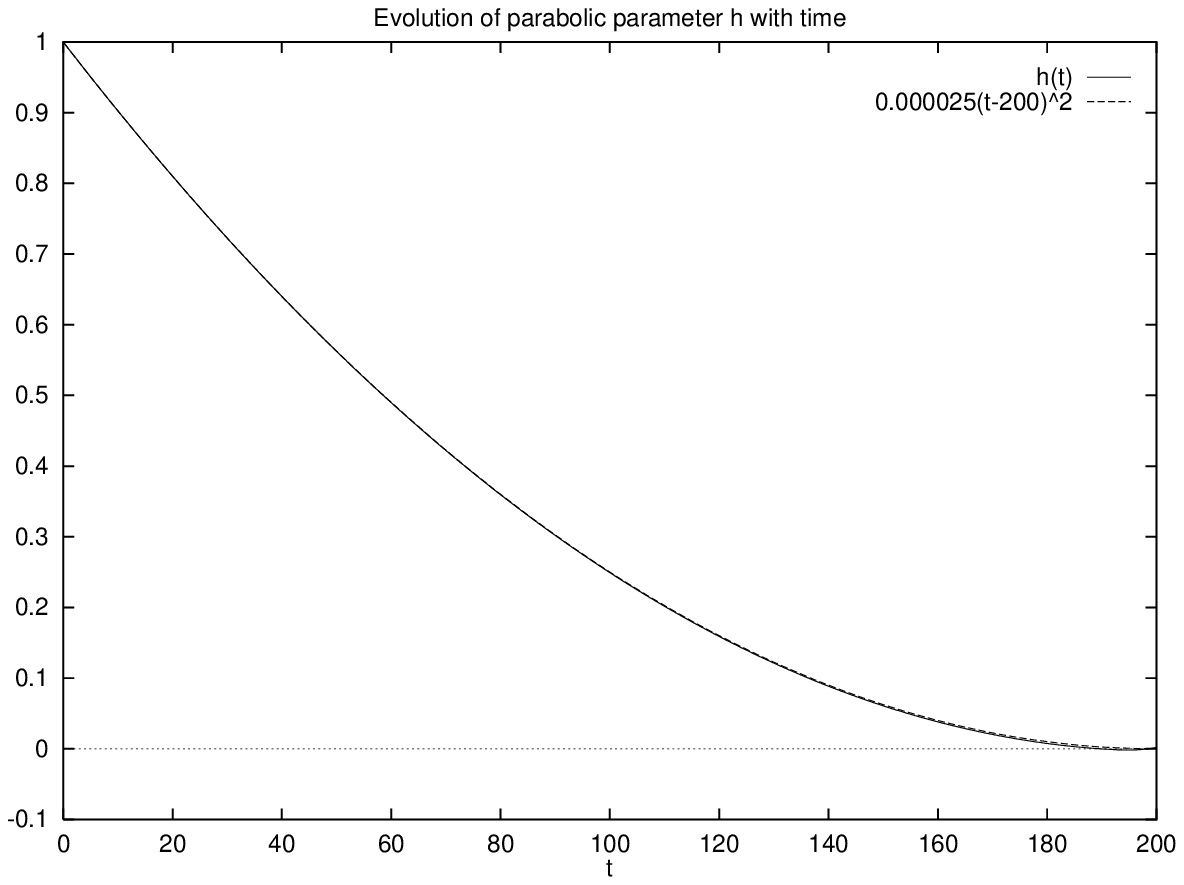}
\caption{(4+1)-dimensional Yang Mills model: Parabolic parameter $h$ as a function of time evolves as $f(0,t)$.}
\label{hp4p1}
\end{center}
\end{figure}

The other parameter in equation (\ref{parabform}) for the parabola,
$h$, should be given by the height of the origin, but this was
calculated in the previous section to be $a(t-T)^2$, substituting the
expressions for $c$ and $T$ we obtain: \[ h(t) =
\frac{v_0^2}{4f_0}\left(t - \frac{2f_0}{|v_0|}\right)^2.\] This is
indeed the correct form, as shown in figure \ref{hp4p1}.  The
initial conditions were $v_0 = -0.01$ and $f_0 = f(0,0) = 1.0$, hence
$h(t) = 0.000025(t - 200)^2$.  The plot of the function overlays the
data.

Now, using both expressions for $p$ and $h$, one can get the general
form of a parabolic $f(r,t)$, which is \begin{equation} f(r,t) =
\frac{v_0^2}{8f_0}r^2 + \frac{v_0^2}{4f_0}\left(t
-\frac{2f_0}{|v_0|}\right)^2\label{pf}\end{equation}
Substitute this into the partial differential equation (\ref{PDEa}),
get a common denominator and simplify to obtain:

\begin{eqnarray*}\lefteqn{ \frac{v_0^2}{4f_0}\left[-\frac{v_0^2}{4f_0}r^2 + 2\frac{v_0^2}{4f_0}\left(t - \frac{2f_0}{|v_0|}\right)^2 + 2r^2\right] \stackrel{?}{=} }\\
& & \frac{v_0^2}{4f_0}\left[\frac{v_0^2}{4f_0}r^2 + 2\frac{v_0^2}{4f_0}\left(t - \frac{2f_0}{|v_0|}\right)^2 + 2r^2\right].\end{eqnarray*}
Clearly these two sides are not the same, and the difference between them is
\[ 2\left(\frac{v_0^2}{4f_0}\right)^2r^2.\]
Since our concern is the adiabatic limit, $v_0^2/(f_0)$ is always
chosen to be small, and this difference goes to zero in the adiabatic
limit.  This term is always much smaller than the $2r^2$ term, and
since we only have accurate numerics away from the neighborhood of the
singularity, the term with $t - 2f_0/|v_0|$ is on the order of
magnitude $f_0/|v_0|$, which is large compared to the correction.
This provides an improved approximate solution for this model which
becomes exact in the adiabatic limit.  This same behavior is
found in section 3.5

The form of the connection $A$ given by this parabolic model is
\[ A(r,t) = \frac{1}{2} \left\{ \frac{\bar{x}dx - d\bar{x}x}
{\left(1 -\frac{v_0^2}{8f_0}\right)r^2 + \frac{v_0^2}{4f_0} \left(t -
\frac{2f_0}{|v_0|}\right)^2}\right\}.\] This is close to the
connection $A$ given by the geodesic approximation multiplied by an
overall factor. This form  suggests a relativistic correction.

\clearpage

\section{The Charge 2 Sector of the $S^2$ Sigma Model}

The first thing to identify in this problem are the static solutions
determined by equation (\ref{genPDEb}).  These are outlined in
\cite{Ward} among others.  The entire space of static solutions can be
broken into finite dimensional manifolds $\mathcal{M}_n$ consisting of
the harmonic maps of degree $n$.  If $n$ is a positive integer, then
$\mathcal{M}_n$ consists of the set of all rational functions of $z =
x + iy$ of degree $n$. We restrict our attention to $\mathcal{M}_2$,
the charge two sector, on which a generic solution has the form
\[ u  = \alpha + \frac{\beta z + \gamma}{z^2 + \delta z + \epsilon} \]
depending on the five complex parameters $\alpha, \beta, \gamma,
\delta, \epsilon$.  To simplify consider only solutions of the form
\[\frac{\gamma}{z^2}\] 
with $\gamma$ real.  The geodesic approximation predicts that
solutions evolve close to 
\[ \frac{\gamma(t)}{z^2}.\]
In this investigation, we use a radially symmetric function $f(r,t) =
\gamma$, and calculate the evolution equation for $f(r,t)$.  It is:

\begin{equation} \fddot = \drrf + \frac{5\drf }{r} - \frac{8r^3\drf }{f^2 + r^4} +
\frac{2f}{f^2 + r^4}\left((\fdot)^2 -
(\drf)^2\right).\label{PDEc}\end{equation} 

We can evaluate the validity of the geodesic approxmation by
evaluating how $f(0,t)$ evolves versus the predicted evolution and how
$f(r,T)$ with $T$ fixed varies from being a horizontal line.

Immediate similarities can be seen between (\ref{PDEc}) and equation
(\ref{PDEa}).  The static solutions are $f(r,T) = c$ for $T$ fixed and
any constant c.  Here the length scale is given by $\sqrt{c}$.  We
investigate progression from $f(r,0) = c_0 > 0$ towards the
singularity at $f(r,T) = 0$.

\subsection{Numerics for the $S^2$ Sigma Model Charge 2 Sector}

As with the other models, a finite difference method is used to
compute the evolution of (\ref{PDEc}) numerically.  Centered
differences are used consistently except for
\[ \drrf + \frac{5\drf }{r}. \]
A naive central differencing scheme here yields serious
instabilities at the origin.  Similar to what we did in section
2.1, let
\[ \mathcal{L} f = r^{-5} \partial_r r^5 \partial r f = \drrf + \frac{5\drf }{r}.\]
This operator has negative real spectrum, hence it is stable.  The
natural differencing scheme for this operator is
\[ \mathcal{L}f \approx r^{-5}\left[ \frac{\left(r + \displaystyle{\frac{\delta}{2}}\right)^5\left(\displaystyle{\frac{f(r+\delta) - f(r)}{\delta}}\right) - \left(r-\displaystyle{\frac{\delta}{2}}\right)^5\left(\displaystyle{\frac{f(r) - f(r-\delta)}{\delta}}\right)}{\delta}\right].\]
This is the differencing scheme used for these terms.

In \cite{LPZ} studying the charge 1 sector of the $S^2$ sigma model it
was found that under their numerical procedure stationary solutions
were unstable and would evolve in time.  This model has no such
problems with the stationary solutions.  They do not evolve in time
unless first bumped with an initial velocity.
 
Now with the differencing explained, we derive $f(r,t+\delt)$ in
exactly the same manner as for the (4+1)-dimensional model in section
2.1 and the charge 1 sector in \cite{Linhart2}.  We have an initial guess for
$f(r,t+\delt)$, either given by $f(r,t+\delt) = f(r,t) + v_0\delt$
with $v_0$ the initial velocity given in the problem, or on subsequent
time steps $f(r,t+\delt) = 2f(r,t) - f(r,t-\delt)$.  We use this guess
to compute $\fdot(r,t)$ on the right hand side of (\ref{PDEc}), and
then we can solve for a new and improved $f(r,t+\delt)$ on the left
hand side of (\ref{PDEc}).  Iterate this procedure to get increasingly
accurate values of $f(r,t)$.

The boundary condition at the origin is found by requiring that
$f(r,t) $ is an even function, hence \[f(0,t) = \frac{4}{3} f(\delr,t)
- \frac{1}{3} f(2\delr, t).\] At the $r = R$ boundary the function
should be horizontal so $f(R,t) = f(R -\delr,t)$.

Subsequent investigation of this model indicated that the appropriate
form of $f(r,t)$ was a parabola instead of a line, and in order to
investigate this phenomenon, the $f(R,t)$ boundary condition was
changed to reflect this.  For the runs with parabolic initial data, we set the boundary condition at $R$ to be:
\[ \drf(R,t) = \drf(R-\delr,t)\frac{R}{R-\delr}.\]

\subsection{Predictions of the Geodesic Approximation for the $S^2$ 
Sigma Model}

Equation (\ref{cp1lag}) gives us the Lagrangian for the general
version of this problem.  We are using
\[ u = \frac{\lambda}{z^2}\]
for our evolution, and via the geodesic approximation we restrict the
Lagrangian integral to this space, to give an effective Lagrangian, as
we did in section 2.2.  The integral of the spatial derivatives of $u$
gives a constant, and hence can be ignored.  Under these assumptions,
up to a multiplicative constant, the effective Lagrangian becomes

\[ L = \int_0^{\infinity} r dr \frac{r^4(\partial_t\lambda)^2}{\left(r^4 + \lambda^2\right)^2}\]
which integrates to
\[ L = \frac{(\partial_t\lambda)^2\pi}{8\lambda}.\]
Since the potential energy is constant, so is the kinetic energy,
hence 
\[ \partial_t\lambda = k \sqrt{\lambda}. \]
Integrating this one obtains
\[ \lambda = (c_1 t + c_2)^2.\]
If $\lambda = 0$ occurs at time $T$, we find 
\[ T = - \frac{c_2}{c_1},\]
hence we rewrite this as
\[ \lambda(t) = a(t-T)^2.\]
This is exactly the same as the evolution predicted in section 2.2 for
the Yang Mills Lagrangian.  Since equation (\ref{PDEc}) tends towards
the linear wave equation when $r\rightarrow\infinity$, the interesting
behavior occurs at the origin.  This is how we predict that
$f(0,t)$ will evolve.

\subsection{Results for the $S^2$ Sigma Model: Evolution of $f(0,t)$}

The computer model was run under the condition that $f(r,0)
= f_0$ with various small velocities.  The initial velocity is
$\partial_t{f}(r,0) = v_0$, other input parameters are $r_{max} = R$,
$\delr$ and $\delt$.

\begin{figure}[H]
\begin{center}
\epsfig{file=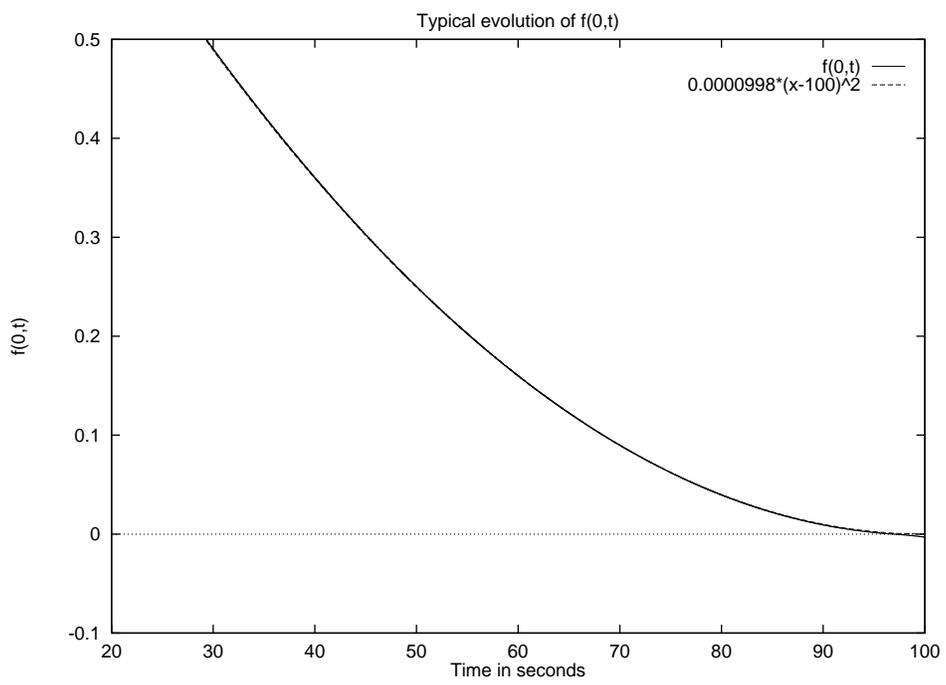}
\caption{$S^2$ sigma model: Evolution of $f(0,t)$ and overlaying fit to parabola.}
\label{fc2}
\end{center}
\end{figure}

A typical evolution of $f(0,t)$ is given in Figure \ref{fc2}.  This is
best modeled by a parabola of the form $a(t-T)^2$, exactly as
predicted.  This curve in particular is approximated by $0.0000998(t -
100.)^2 $.

We fit $f(0,t)$ to a parabola of the form $a(t-T)^2$ for various
initial conditions.  Table \ref{fch2} gives initial conditions $f_0
= f(r,0)$ and $v_0 = \partial_t{f}(r,0)$ and the subsequent $T$ and $a$ when
$\delr = 0.025$ and $\delt = 0.001$.

Exactly as in section 2.3, we have
\begin{eqnarray*} T &=& \frac{2f_0}{|v_0|}\\
\noalign{\hbox{and}}\\
a &=& \frac{v_0^2}{4f_0}.
\end{eqnarray*}
\begin{table}[h]
\begin{center}
\caption{$S^2$ sigma model: Parameters for best fit parabola to $f(0,t)$ vs. initial data $f_0$ and $v_0$.} 
\[\begin{array}{*{3}{r}{l}}  f_0 &  v_0\ &\ \ \ T\ \ & a\ \\ 
1.0&    -0.01&   200&   0.0000250\\
1.0&    -0.02&   100&   0.0000998\\
1.0&    -0.03&    67&   0.000224\\
1.0&    -0.04&    50&   0.000398\\
0.5&    -0.01&   100&   0.0000499\\
2.0&    -0.01&   400&   0.0000125\\
3.0&    -0.01&   600&   0.00000833\\
\end{array}\]
\label{fch2}
\end{center}
\end{table}

\clearpage

\subsection{Characterization of Time Slices $f(r,T)$: Evolution of a Horizontal Line in the $S^2$ Sigma Model}

\begin{figure}[h]
\begin{center}
 \epsfig{file=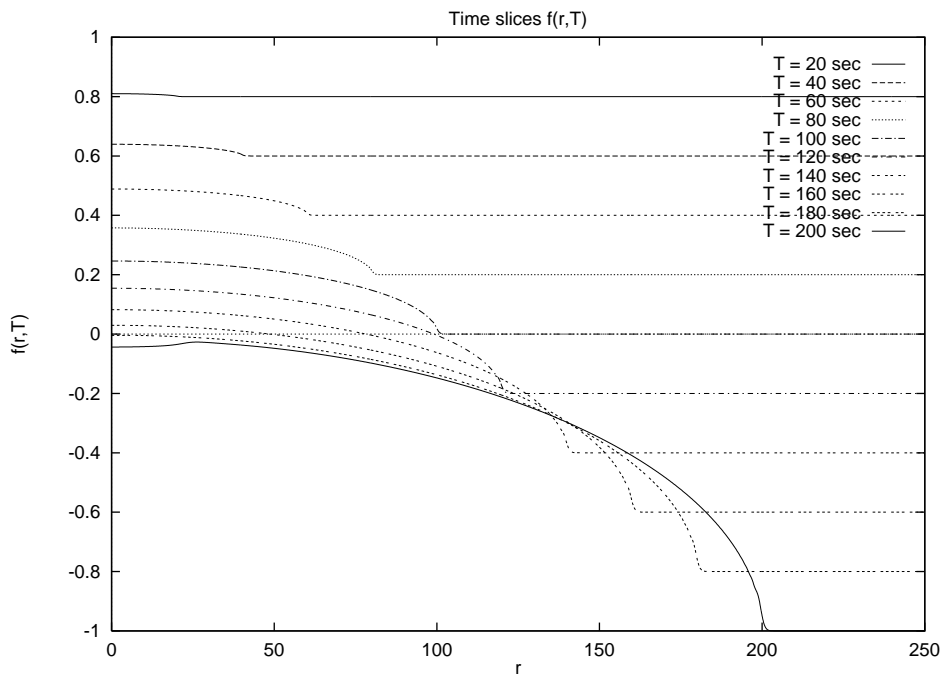}
\caption{$S^2$ sigma model: Time slices of $f(r,T)$  evolve an elliptical bump at the origin.}
\label{tsch2}
\end{center}
\end{figure}

With the evolution of $f(0,t)$ taken care of, we consider the shape of
the time slices $f(r,T)$ for a given fixed $T$.  As in section 2.4
with the (4+1)-dimensional Yang Mills model, this is rather striking.
An elliptical bump forms at the origin, as seen in figure \ref{tsch2}.

Once again exactly as in section 2.4 with the (4+1)-dimensional model,
this elliptical bump has equation
\begin{equation} \frac{x^2}{a^2} + \frac{(y-k)^2}{b^2} = 1.\label{ellbumpch2}
 \end{equation}
The question naturally arises how the parameters $a$, $b$, and $k$ evolve, and as in section 2.4, we once again have
\begin{eqnarray*} a &=& t \\
 b &=& \frac{v_0^2}{4f_0} t^2\\
 k &=& f_0 + v_0t 
\end{eqnarray*}

Close to the origin the ellipse is well approximated by a horizontal line,
giving further credence to the validity of the geodesic approximation.

\clearpage

\subsection{Characterization of Time Slices $f(r,T)$: Evolution of a Parabola
in the $S^2$ Sigma Model}

As with the (4+1)-dimensional model in section 2.5, the evolution of the ellipse
suggested the curve was trying to obtain the shape of a parabola of the form

\begin{equation} f(r,t) = p r^2 + h. \label{parabch2}\end{equation}

To get the general form of the parabola, we follow the calculation
from the (4+1)-dimensional model in section 2.5. From our ellipse equation
(\ref{ellbumpch2}):
\[ \frac{dy}{dx} = -\frac{x^2b^2}{(y-k)a^2}\]
so
\[ \frac{d^2y}{dx} = \frac{-b^2}{(y-k)a^2} - \frac{xb^2}{(y-k)^2a^2}\frac{dy}{dx}\]
At $x = 0$, $y-k = b$ and this gives
\[ \frac{d^2y}{dx} = \frac{-b}{a^2}.\]
Recall from the previous section that $b = ct^2$ and $a = t$, so this
gives \[ \frac{d^2y}{dx^2} = -c.\] The identification of $c$ gives \[
\frac{d^2y}{dx^2} = -\frac{v_0^2}{4f_0}.\] So \[p =
-\frac{1}{2}\frac{d^2y}{dx^2} = -\frac{v_0^2}{8f_0}.\]  Rather unsurprisingly, this echoes the result of section 2.5. 

\begin{figure}[H]
\begin{center}
\epsfig{file=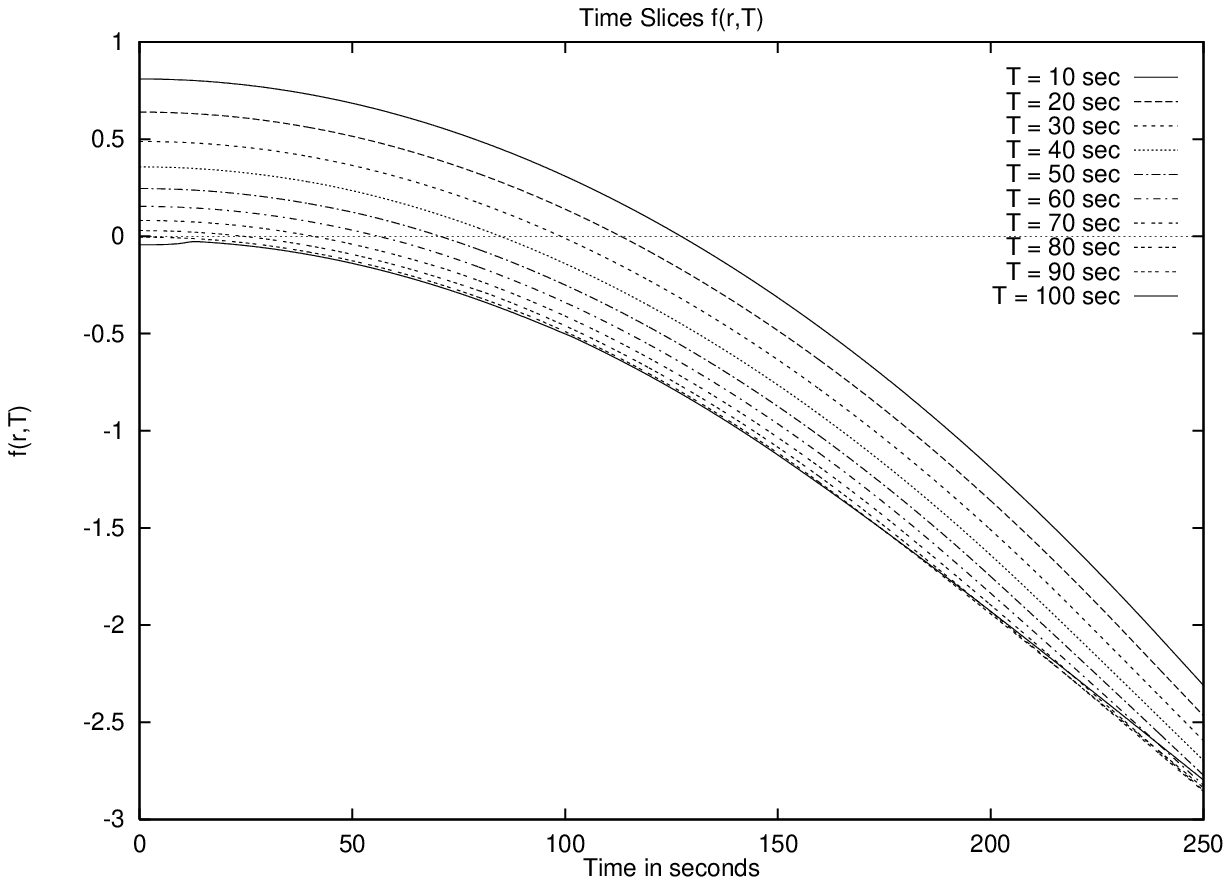}
\caption{$S^2$ sigma model: Time slices of the evolution of a 
parabola
are parabolas.}
\label{pch2}
\end{center}
\end{figure}

\begin{figure}[H]
\begin{center}
\epsfig{file=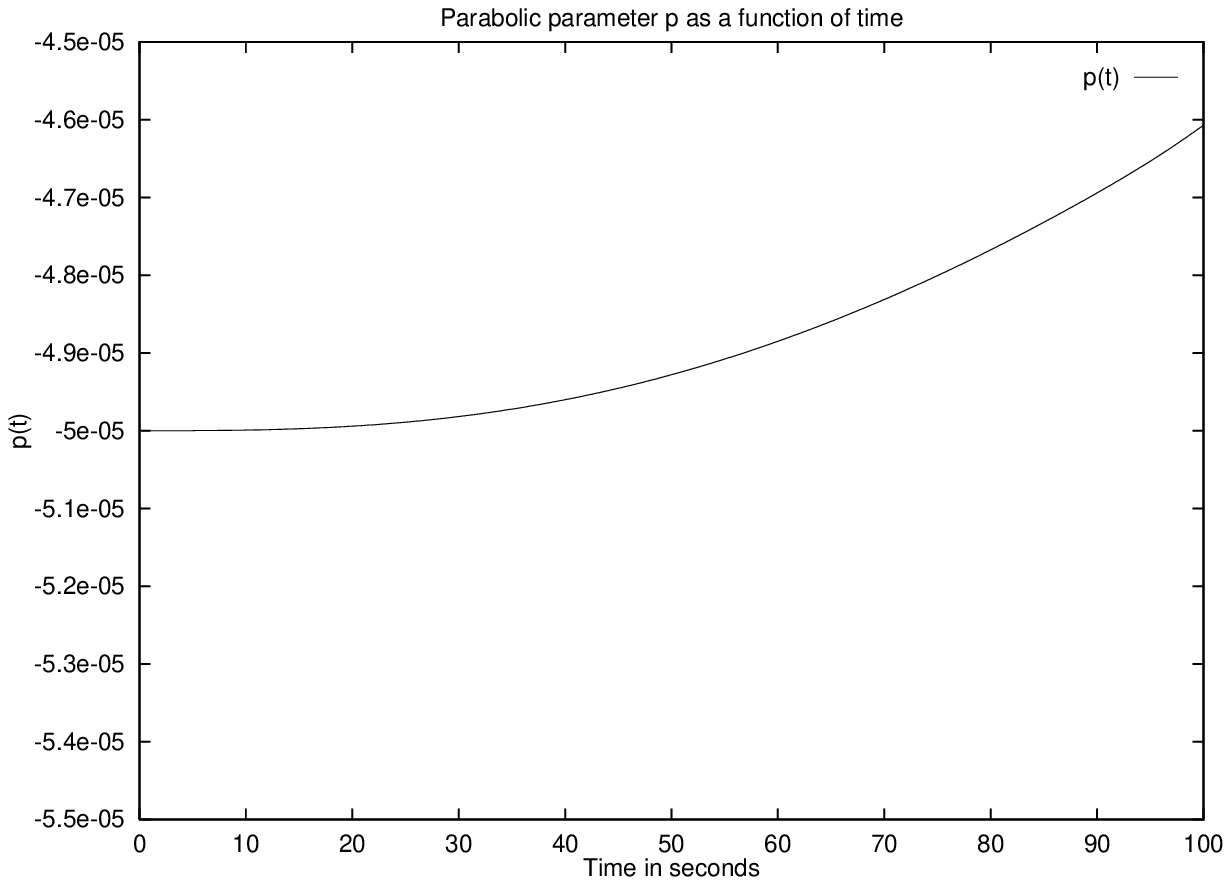}
\caption{$S^2$ sigma model: Evolution of parabolic parameter $p$ with time.}
\label{ptch2}
\end{center}
\end{figure}

\begin{figure}[H]
\begin{center}
\epsfig{file=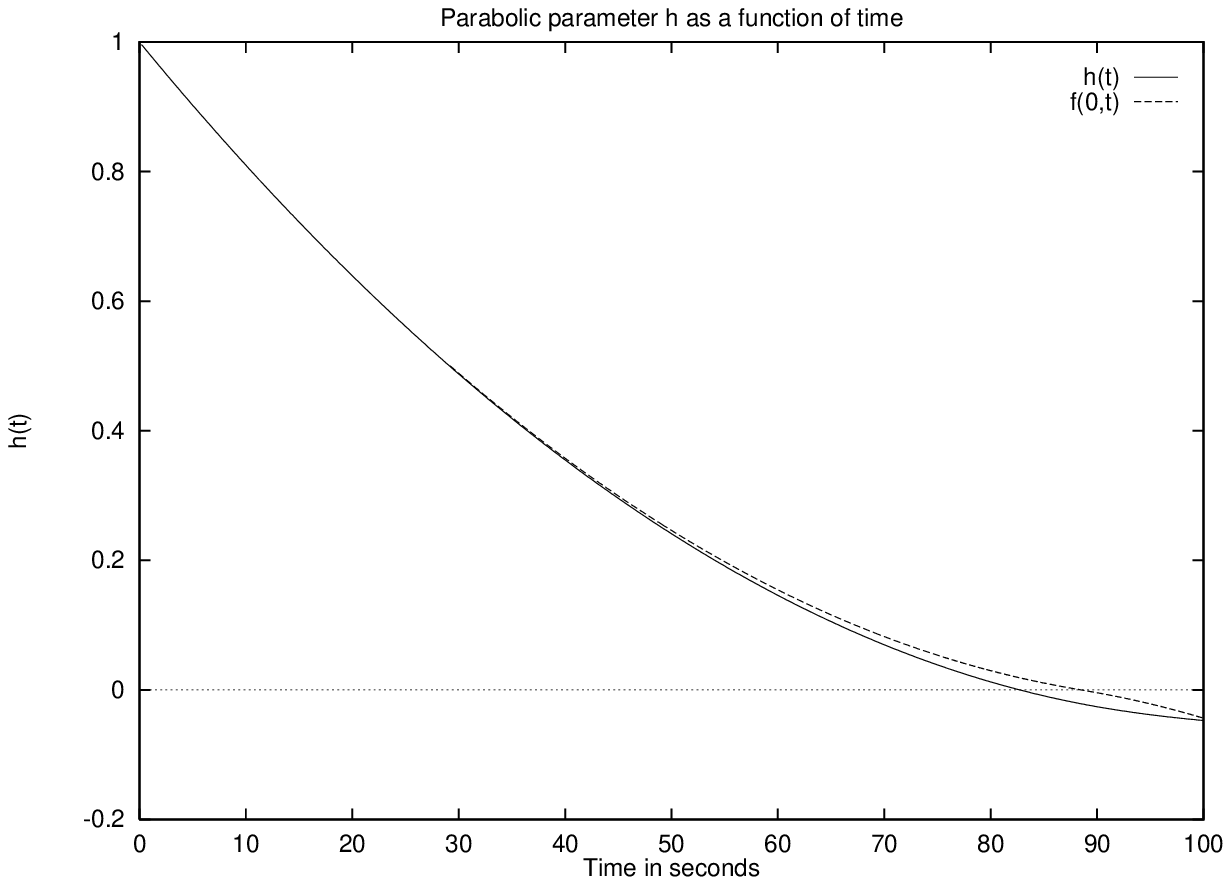}
\caption{$S^2$ sigma model: Evolution of parabolic parameter $h$ with time, comparison to $f(0,t)$.}
\label{htch2}
\end{center}
\end{figure}

When a run is started with this initial data, $\partial_t{f_0} = v_0 =
-0.02$, $f_0 = f(0,0) = 1.0$ and $p = -\frac{v_0^2}{8f_0} = -0.00005$,
the time slices of the data have this same profile.  This is shown in
figure \ref{pch2}.  The parabolic parameter $p$ varies by less than
1 part in 10 during the run as seen in figure \ref{ptch2}.  The
parabolic parameter $h$ evolves close to $f(0,t)$ as seen in figure
\ref{htch2}.

For a parabola of the form
\[ f(r,t) = p(t)r^2 + h(t), \]
we have 
\[ p(t) = -\frac{v_0^2}{8f_0},\]
{\it i.e.\ } $p(t)$ is constant, and
\[ h(t) = \frac{v_0^2}{4f_0}\left(t - \frac{2f_0}{|v_0|}\right). \]

Using the identification of $p$ and $h$ in the parabolic form of
$f(r,t)$ and letting
\[ \tau = t - \frac{2f_0}{|v_0|}\]
we have
\[ f(r,t) = -\frac{v_0^2}{8f_0}r^2 + \frac{v_0^2}{4f_0}\tau^2. \]
Substitute this into the partial differential equation (\ref{PDEc}), get a common denominator, and simplify to obtain
\begin{eqnarray*}\lefteqn{ \frac{v_0^6}{32 f_0^3}\left(\frac{r^4}{4} - r^2\tau^2 + \tau^4\right) + \frac{v_0^2}{2f_0}r^4 \stackrel{?}{=}}\\
& & -\frac{v_0^6}{32f_0^3}\frac{r^4}{4} + \frac{v_0^6}{32 f_0^3} \tau^4
+ \frac{v_0^2}{2f_0}r^4.\end{eqnarray*}
The difference between the two sides is
\[ \frac{v_0^6}{64f_0^3}r^4 - \frac{v_0^6}{32f_0^3}r^2\tau^2. \]
As in section 2.5 for the (4+1)-dimensional Yang Mills model, our
concern is with the geodesic approximation, and so $v_0^2/(f_0)$ is
always chosen to be small.  The correction is then much smaller than
the term
\[ \frac{v_0^2}{2f_0}r^4. \]

\clearpage
\section{Conclusions}

The formation of singularities in the geodesic approximation to the
(4+1)-dimensional Yang Mills model and the (2+1)-dimensional $S^2$
sigma model has been studied numerically using radially symmetric
solutions and an iterated finite differencing scheme.

The predictions of the Lagrangians of the two models are that the
trajectory towards blow-up should occur parabolically as $a(t-T)^2$
where $T$ represents the blow-up time.  This is confirmed in the
behavior of both numerical models.

The geodesic approximation in both models predicts that the solution
will evolve in time from horizontal line to horizontal line.  To first
approximation, near the origin, this is what occurs.  A more precise
characterization is available.  Both of these models, when started
with a horizontal line as an initial condition, evolve an elliptical
bump at the origin.

The elliptical bumps suggested that the model preferred a parabolic
initial condition and a parabolic state for the fixed time profile of
the evolution.  

These parabolic solutions become exact in the adiabatic limit, and
provide alternative approximate solutions to the differential
equations.

\section{Acknowledgments}

I would like to thank my dissertation supervisor, Lorenzo Sadun, for
his constructive comments and suggesstions for additional lines of
research and improvement on this manuscript and my dissertation.

\bibliography{art}
\end{document}